\documentclass[aps ]{revtex4}
\usepackage{epsfig,graphics,amssymb,amsmath}
\def\Re{{\rm Re}}\def\ex{{\bf e}_x}\def\ey{{\bf e}_y}\def\ez{{\bf e}_z}
\def\bO{\boldsymbol{\Omega}}\def\U{{\bf U}}
\def\Vperp{V_\perp}
\def\Reg{\Re_{\dot{\gamma}}}

\usepackage{eso-pic}
\usepackage{color}
\usepackage{type1cm}
\makeatletter
  \AddToShipoutPicture{%
    \setlength{\@tempdimb}{.5\paperwidth}%
    \setlength{\@tempdimc}{.5\paperheight}%
    \setlength{\unitlength}{1pt}%
    \put(\strip@pt\@tempdimb,\strip@pt\@tempdimc){%
    }
}
\makeatother

\begin{document}
\title{Continuous breakdown of Purcell's scallop theorem with inertia}
\author{Eric Lauga\footnote{Email: lauga@mit.edu}}
\affiliation{Department of Mathematics,
Massachusetts Institute of Technology,
77 Massachusetts Avenue,
Cambridge, MA 02139.}
\date{\today}
\begin{abstract}

Purcell's scallop theorem defines the type of motions of a solid body - reciprocal motions -  which cannot propel the body in a viscous fluid with zero Reynolds number.  For example, the flapping of a wing is reciprocal and, as was recently shown,  can lead to directed motion only if its frequency Reynolds number, $\Re_f$, is above a critical value of order one. Using elementary examples, we show the existence of oscillatory reciprocal motions which are effective  for all arbitrarily small values of the frequency Reynolds number and induce net velocities scaling as $\Re_f^\alpha$ ($\alpha > 0$). This  demonstrates a continuous  breakdown of the scallop theorem with inertia.

\end{abstract}
\maketitle

A large variety of biological movements occur in a fluid environment, from swimming bacteria to   whales. In many cases, the study of fluid forces is crucial to the understanding of animal locomotion \cite{gray68,lighthill76,brennen77,purcell77,childress81,alexander02}. Because of the large range of relevant length scales in biological motility - eight orders of magnitude in size, from less than a hundred nanometers to tens of meters - fluid mechanics occurs in distinct regimes with important mechanical consequences. On small length scales, the relevant Reynolds number is usually very small ($\Re\approx 10^{-4}$ for swimming {\it E. coli}) and viscous forces are dominant. This is the Stokesian realm of swimming microorganisms such as bacteria, spermatozoa, and ciliated cells. At the opposite end of the range of length scales, the Reynolds numbers are typically very large ($\Re\approx 10^7$ for a swimming tuna) and inertial forces are dominant. This is the Eulerian realm of flying birds and swimming fishes.  In this paper, we address the transition from the Stokesian to the Eulerian realm, and show that, in some situations, this transition can take place continuously with an increase of the relevant Reynolds number.

In his 1977 lecture {\it Life at low Reynolds numbers}, Edward Purcell introduced the ``scallop theorem" \cite{purcell77}. He observed that the Stokes equations, which govern fluid flows at zero Reynolds numbers and are both linear and independent of time, are identical under time reversal. Consequently, there exists a certain geometrical class of motion (or, more generally, actuation of a solid body), termed ``reciprocal motion'', which cannot lead to any locomotion in this limit.  A reciprocal motion (or actuation) is a motion in which the geometrical paths followed by various material points on the body are identical when viewed under time reversal. By symmetry, such motion can only lead to a net movement equal to minus itself, and therefore, no net movement at all (see also Refs. \cite{childress81,childress04}). The simplest example of a reciprocal motion is a periodic motion composed of two distinct parts. In the first part, the body moves in a certain prescribed way, and in the second part, the body moves in a manner which is identical to the first-part as seen under time reversal. A scallop opening and closing belongs to this subclass of reciprocal motion and, independently of the rate of opening and closing, the scallop cannot move.

Another example of reciprocal motion - or, in this case, reciprocal actuation - is a flapping body. Consider a solid body oscillated up and down in translation in a prescribed manner by an external means. Since the motion going up is the time-reversal symmetry of the motion going down, the flapping body does not move on average in the limit of zero Reynolds numbers. However, large animals such as birds use flapping wings for locomotion, and so clearly a thin flapping body must be effective in the Eulerian realm.  The question then arises: When does a flapping body - or more generally a reciprocal motion -  become effective? How much inertial force is necessary to break the constraints of the scallop theorem?

This question was first formulated and studied by Childress and Dudley  \cite{childress04}. The  mollusc {\it Clione antarctica} was observed to possess two modes of locomotion. The first is non-reciprocal and uses  cilia distributed along the body of the mollusc. The second  is reciprocal and consists of two flapping wings. The flapping-wing mode was observed to be predominant for the large  swimming velocities. Using both experimental observations and fluid mechanics models, the authors postulated that reciprocal motions are ineffective in producing any net motion unless the relevant frequency - or ``flapping'' - Reynolds number, $\Re_f$, is sufficiently large (order unity). In other words, the transition from no-motion to motion occurs at a finite value of $\Re_f$ and the breakdown of the scallop theorem is discontinuous. This idea was subsequently studied in laboratory experiments \cite{vandenberghe04,vandenberghe06} and numerical simulations \cite{alben05,lu06} of flapping symmetric bodies, both of  which confirmed the transition to directed motion as a symmetry-breaking instability occurring at a finite value of the frequency Reynolds number, as well as the robustness of this transition to a change in a variety of geometrical and mechanical parameters.

In this paper, we consider a series of elementary  oscillatory reciprocal motions of a solid body with broken spatial symmetries and show that they become effective in producing a net translation of the body  for arbitrarily small values of the frequency  Reynolds number, with induced velocities scaling  as $\Re_f^\alpha$ ($\alpha > 0$, {\it inertial creep} \cite{childress_conf}). This demonstrates a continuous breakdown of the scallop theorem with inertia.

The examples we propose rely on classical results of lift forces for the motion of spherical particles at  small Reynolds number \cite{proudman57,segre62a,segre62b,leal80}. We consider a solid spherical particle (density $\rho_p$, radius $a$) oscillating with frequency $\omega$ and amplitude $d$  in a fluid of density $\rho$ and shear viscosity $\mu$. The three different setups we propose are described below, and we start by some general remarks. In the case of purely translational motion, including the effect of inertia on the particle motion can be done in a number of limits, as there are in general three relevant Reynolds numbers. Firstly, the unsteady term in the Navier-Stokes equations scales as $ \rho \omega U_0$ (where $U_0 = d\omega$ is the typical speed of translation), and is smaller than the  typical viscous term, of order $\mu U_0/a^2$, by a factor of $\Re_\omega= a^2 \omega/\nu$, where $\nu=\mu/\rho$ is the kinematic viscosity. 
Secondly, the nonlinear advective term in the Navier-Stokes equations scales with $ \rho U_0^2/a$, and is smaller than the  viscous term by a factor of $\Re_f= a U_0/\nu= a d\omega /\nu$, which is the flapping (or ``frequency'') Reynolds number \cite{vandenberghe04,vandenberghe06,alben05}.  
Thirdly, the particle inertia is quantified by a particle Reynolds number,  $\Re_p = \rho_p a^2 \omega / \mu$, the ratio of the typical  rate of change of the particle momentum,  $\rho_p a^3 U_0 \omega$, to the typical viscous forces on the particle, $\mu a U_0$. In this paper, we will consider  the asymptotic limit where
\begin{equation}\label{limit}
\{\Re_p,\Re_\omega\}\ll \Re_f \ll 1,
\end{equation}
so that the motion of the flapper is quasi-static and the leading-order departure of the fluid forces from the Stokes laws is due to the nonlinear advective term in the Navier-Stokes equations  \cite{footnote1}. 
The limit described by Eq.~\eqref{limit} is equivalent to that of small frequency Reynolds number ($\Re_f \ll 1$) and large flapping amplitude ($a/d \ll 1$ and $a/d \ll \rho/\rho_p$). Note that  this is a different limit from the work in Refs. \cite{vandenberghe04,alben05,vandenberghe06,lu06} where body inertia likely played an important role. We consider below three examples of such large-amplitude low-$\Re_f$ reciprocal flapping  which lead to directed motion for arbitrarily small values of $\Re_f $.

\begin{figure}[t]
\begin{center}
\includegraphics[width=.95\textwidth]{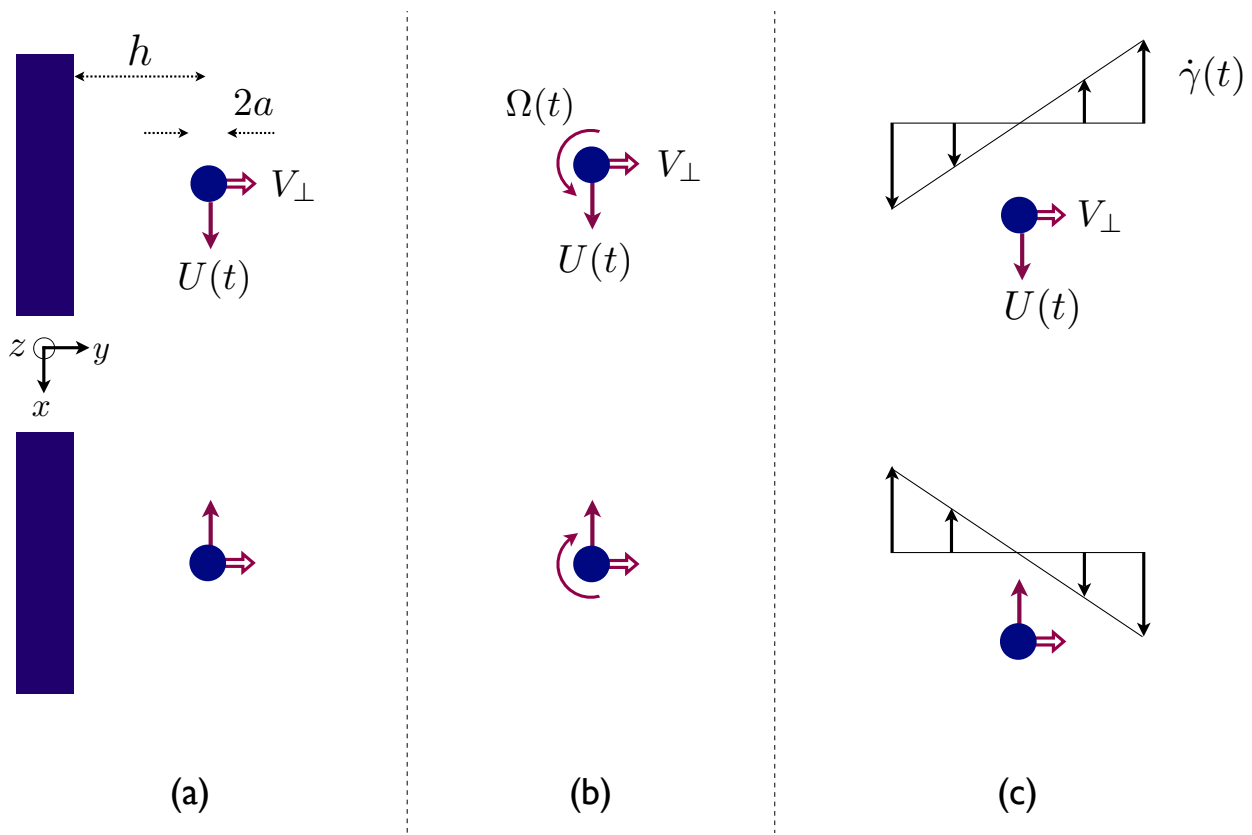}
\caption{Three examples of reciprocal forcing leading to translation of a solid body (sphere of radius $a$) for arbitrarily small values of the frequency Reynolds number. In each example, the reciprocal motion is composed of the periodic repetition of two distinct parts, with the second part (bottom) being identical to the first part (top) as seen under time reversal: (a) Oscillation in vertical position of a sphere parallel to a solid surface leads to motion perpendicular to the surface;
(b) In-phase oscillations in translation and rotation of a sphere leads to motion perpendicular to both the directions of translation and rotation;
(c) Oscillation in vertical position of a sphere in an oscillating shear flow (in phase) leads to motion perpendicular to the direction of translation;
In all cases,  $\Vperp$ denotes the (small) sphere velocity induced by inertial forces. In case (a), the distance to the solid surface is denoted $h$.
}
\label{mainfig}
\end{center}
\end{figure}

The first example is that of a flapper near a wall. Specifically, we consider the reciprocal oscillation in vertical position of the solid sphere with velocity ${\bf U}(t) = U(t) \ex$ parallel to a stationary solid surface and free to move in the $y$ and $z$ directions (see notations in Fig.~\ref{mainfig}a).  
In the Stokes flow limit ($\Re_f=0$), the sphere experiences no lift force, and  remains at a constant distance, $h$, to the solid surface.  The first effect  of inertia on this problem, in the limit set by Eq.~\eqref{limit},  is the appearance of a lift force, directed away from the solid surface, and independent of the sign of  ${ U(t)}$ \cite{cox77,vasseur77,leal80}. Such a limit is captured when the Oseen length scale $\nu/U_0$, the distance away from the sphere where inertial forces become important, is much larger than all relevant length scales of the problem, {\it i.e.} the sphere radius, $a$, and its distance to the surface, $h$. In the simple case where  $a \ll h \ll \nu/U_0$, the lift force leads to a low-Reynolds number lift velocity for the particle \cite{cox77,vasseur77,leal80}
\begin{equation}\label{Vperp_a}
{\bf V}_\perp (t)= \Vperp (t) \ey,\quad \Vperp(t) = \frac{3}{32}\frac{aU(t)^2}{\nu},
\end{equation}
always directed away from the surface.
For an oscillatory motion,   $U(t) = U_0 \cos \omega t$,  the lift velocity away from the surface averages over one period to
\begin{equation}\label{dperp_a}
\frac{\langle V_\perp \rangle  }{U_0}
= \frac{3}{64} \Re_f. 
\end{equation}
A flapper near a wall performing a reciprocal translational motion is therefore able to move forward (away from the wall) for arbitrarily small values of the frequency Reynolds number. This inertial migration decreases to zero with the first power of the Reynolds number ($\alpha =1$), and the Stokes limit is recovered when we formally  set  $\Re_f=0$ in Eq.~\eqref{dperp_a}.

Our second example is that of a rotating flapper. We consider the case where the solid sphere is oscillating  both in translation and rotation, with velocity and rotation rates given by ${\bf U} (t) =U(t) \ex$ and $\bO(t) =\Omega(t) \ez$, and is free to move in the $y$ and $z$ direction (see  Fig.~\ref{mainfig}b). If the two oscillations are in phase, the actuation of the sphere is reciprocal, which we will assume here, and no average motion is obtained in the Stokes limit. If $\Omega_0 $ is the typical magnitude of $\bO(t)$, the rotation Reynolds number $\Re_\Omega=a^2 \Omega_0 /\nu$ measures the importance of inertial forces due to the rotational motion.   In the asymptotic limit set by Eq.~\eqref{limit}, and for $\Re_\Omega \sim \Re_f $, the  first effect of inertia  is the appearance of a lift force perpendicular to both the directions of translation and rotation  \cite{rubinow61,cox65,leal80} and given by ${\bf F}_L=\pi a^3 \rho \bO \times \U$. This results in  a low-Reynolds number lift velocity
\begin{equation}
{\bf V}_\perp (t)= \Vperp (t)\ey,\quad\Vperp (t)
= \frac{a^2 U(t) \Omega(t)}{6\nu} \cdot
\end{equation}
When $U(t) =  a\Omega(t) = U_0 \cos \omega t$, we obtain an average translational velocity, along the $y$ direction, given by 
\begin{equation}
\frac{\langle V_\perp \rangle }{U_0}
= \frac{ \Re_f}{12}  .
\end{equation}
Here again, the reciprocal translational  and rotational motion of the solid sphere leads to a directed motion for  arbitrarily small values of the Reynolds number. The magnitude of this directed motion also decreases to zero with the first power of $\Re_f$ ($\alpha=1$). 

As a final example, we show that these results are also valid when the fluid in the far-field is not quiescent by considering a flapper in a shear flow. Specifically, as shown in Fig.~\ref{mainfig}c, we consider the case when the  solid sphere  is oscillating in vertical position with a prescribed velocity, ${\bf U}(t) = U(t)\ex$, in a shear flow described by the  far-field undisturbed flow field  ${\bf u}_\infty  = - \dot{\gamma}(t) y \ex$ (the center of the sphere is located at $y=0$) and is free to move in the $y$ and $z$ directions. If the two oscillations are in phase, the motion of the sphere is reciprocal, which we assume here, and no average motion is obtained in the limit of zero Reynolds number. We also assume that the sphere is far away from the surfaces responsible for the creation of the shear flow and therefore ignore wall effects \cite{ho74,vasseur76,leal80}.  If $\dot{\gamma}_0$ denotes the typical magnitude of $\dot\gamma (t)$, an additional Reynolds number, $\Reg=a^2 \dot{\gamma}_0 /\nu$, needs to be introduced.  Here, the first effect of inertia is the appearance of a lift force directed across the undisturbed streamlines  \cite{saffman65,leal80,stone00}. The original study, due to Saffman \cite{saffman65,stone00}, calculated this lift force in the limit  where $ \Re_f \ll \Reg^{1/2} \ll 1$, and in this case the lift force is moving the sphere in the direction opposite to its translational velocity. We consider here the same asymptotic limit, together with the limit assumed in Eq.~\eqref{limit}. In this case, and if $U(t)\cdot \dot{\gamma}(t) > 0$, the sphere experiences a low-Reynolds number lift velocity given by 
\begin{equation}
{\bf V}_\perp (t)= \Vperp (t)\ey ,\quad
\Vperp (t) = c_1 |U(t)| \left( \frac{a^2|\dot{\gamma}(t)|}{\nu}\right)^{1/2},
\end{equation}
where $c_1 \approx 0.343$ is a numerical coefficient. 
For an oscillatory motion $U(t)  = U_0 \cos \omega t$, and with $  \dot{\gamma}(t) = U(t)/a$ to satisfy Saffman's asymptotic limit, we get an average velocity, along the $y$ direction, given by 
\begin{equation}
\frac{\langle V_\perp \rangle }{U_0}
= c_2\Re_f  ^{1/2},
\end{equation}
where $c_2=2c_1\int_0^{\pi/2} (\cos t)^{3/2}\,{\rm d}t /\pi \approx 0.191$.
As in the previous cases, the actuation of the sphere is reciprocal and yet it leads to a directed motion for  arbitrarily small values of the frequency Reynolds number $\Re_f$. Here, however, the magnitude of the induced velocity decreases to zero with the square-root of the Reynolds number ($\alpha=1/2$). Also, in this case, the motion will continue until the point along the $y$ axis where the local velocity from the shear flow cancels out the translational velocity of the sphere.

As a summary, we have presented elementary examples of oscillatory reciprocal forcing of a solid body  leading to net translational motion of the body for arbitrarily small values of the frequency Reynolds number, $\Re_f$. When the frequency Reynolds number is formally set to zero, the effect disappears as dictated by the scallop theorem, but it  remains non-zero for all non-zero values of $\Re_f$.  The induced average velocities scale  as $\Re_f^\alpha$  ($\alpha > 0$), corresponding to  the limit of asymptotically large Strouhal number, ${\rm St} = \omega d /\langle  \Vperp \rangle  \sim  \Re_f^{-\alpha}$. This demonstrates  that the breakdown of Purcell's scallop theorem with inertia can take place in a continuous way without a finite onset of translational motion.

As our examples show, a directed motion on the order of the flapping velocity will take place when $\Re_f\sim 1$.  Moreover, the mechanical efficiencies of the examples above - ratio of the useful work to the total work done by the flapper -  scale as $\Re_f^{2\alpha}$ so that order one efficiencies should also be expected for order one Reynolds numbers. From a biological perspective, both these observations suggest that reciprocal gaits are very inefficient for small Reynolds number and become advantageous only when $\Re_f\sim 1$. Consequently, and even in the absence of a mathematical bifurcation, the onset of an appropriately defined ``efficient flapping flight'' is expected occur at a finite value of  $\Re_f$ \cite{childress04}.

Furthermore, it is important to note that all of our examples  display some  spatial broken symmetries which govern the direction of the net motion of the solid body: (a) the location of the wall, (b) the direction of the rotation rate, and (c) the direction of the shear flow. This is somewhat  different from the ``flapping wing'' setup studied  experimentally in Refs.~\cite{vandenberghe04,vandenberghe06} and numerically in Refs.~\cite{alben05,lu06} where both the shape and the actuation of the wing are symmetric and where locomotion is a result of a hydrodynamic instability \cite{footnote2}.

Finally, we have considered  examples leading to net translational motion, but similar examples exploiting lift forces and torques on asymmetric particles \cite{brenner61,chester62,brenner63,cox65} could be devised leading to a net rotation, or combined translation and rotation, of the solid body \cite{footnote3}.

\section*{Acknowledgments}
We thank H. Chen, A. E. Hosoi, C. Pipe, M. Roper and H. Stone for useful discussions. This work was supported in part by the Charles Reed Fund at MIT and by the National Science Foundation (CTS-0624830)


\begin{thebibliography}{10}
\bibitem{footnote1}
In the biologically relevant situations where $\rho\sim\rho_p$, we have  $\Re_p\sim \Re_\omega$.
\bibitem{footnote2}
Note however that the flapping of an asymmetric  wing was also considered in Ref.~\cite{vandenberghe06} with little influence on the nature of the onset of directed flapping motion.
\bibitem{footnote3}
In a related study, M. Roper and H. A. Stone (personal communication) have recently shown that the reciprocal oscillatory translation of a solid body could lead to a net rotation for arbitrarily small values of the frequency Reynolds number.
\end{thebibliography}
\end{document}